\documentclass[pra,superscriptaddress,twocolumn]{revtex4-1}
\usepackage{xcolor}
\usepackage{graphicx}
\usepackage{amssymb, amsmath, amsthm}
\usepackage{bbold}
\usepackage{dsfont}
\usepackage[percent]{overpic}

\usepackage[colorlinks]{hyperref}

\usepackage[authormarkup=none]{changes}
\definechangesauthor[name={a}, color=orange]{a} 
\definechangesauthor[name={s}, color=blue]{s} 
\definechangesauthor[name={j}, color=green]{j} 

\newcommand{\avg}[1]{\langle #1 \rangle}

\newcommand{\cL}{\mathcal{L}}

\newcommand{\rmd}{\mathrm{d}}

\begin{document}
	
\author{Alexandre Roulet}
\affiliation{Department of Physics, University of Basel, Klingelbergstrasse 82, CH-4056 Basel, Switzerland}
\affiliation{Centre for Quantum Technologies, National University of Singapore, 3 Science Drive 2, Singapore 117543, Singapore}

\author{Stefan Nimmrichter}
\affiliation{Centre for Quantum Technologies, National University of Singapore, 3 Science Drive 2, Singapore 117543, Singapore}

\author{Jacob M. Taylor}
\affiliation{Joint Quantum Institute/NIST, College Park, Maryland 20742, USA}
\affiliation{Joint Center for Quantum Information and Computer Science, University of Maryland, College Park, Maryland 20742, USA}

\title{An autonomous single-piston engine with a quantum rotor}
\begin{abstract}
Pistons are elementary components of a wide variety of thermal engines, allowing to convert input fuel into rotational motion. Here, we propose a single-piston engine where the rotational degree of freedom is effectively realized by the flux of a Josephson loop -- a quantum rotor -- while the working volume corresponds to the effective length of a superconducting resonator. Our autonomous design implements a Carnot cycle, relies solely on standard thermal baths and can be implemented with circuit quantum electrodynamics. We demonstrate how the engine is able to extract a net positive work via its built-in synchronicity using a filter cavity as an effective valve, eliminating the need for external control.
\end{abstract}
\maketitle
\section{Introduction} 
Recent progress in the miniaturization of heat machines has allowed to experimentally explore the realm of quantum thermodynamics, where rules beyond those of its classical counterpart govern the operation. Examples of such remarkable machines are an absorption refrigerator operating at the level of a few phonons~\cite{ions17}, a nano-beam engine fueled with a squeezed non-equilibrium reservoir~\cite{klaers17}, and a nuclear magnetic resonance setup reversing the thermodynamic arrow of time using quantum correlations~\cite{micadei17}. In these implementations, genuine quantum effects could be observed and the benefits of additional quantum resources besides standard thermal reservoirs could be assessed. Yet in order to quantify the actual advantage of using quantum resources, their energetic cost must be included into the thermodynamic balance \cite{scully03,niedenzu16,brandner17,clivaz17}. In the end, does it pay off spending additional energy to implement, say, a squeezed thermal bath to increase the performance of a thermal machine?

A way to address this open question is to design autonomous thermal machines, which draw their energy exclusively from standard thermal baths and do not require the action of an external (quantum) agent to run the cycle~\cite{brask15,silva16}. Inspired by such an engine studied recently~\cite{rotor17}, we propose here a flux-based piston that obeys this autonomous design and can be implemented with circuit quantum electrodynamics (QED) elements. This platform promises timely experimental demonstrations of quantum heat machines~\cite{pekola15}; in fact, several engines have been designed,  for instance with Cooper pairs tunneling across a Josephson junction against a voltage bias~\cite{hofer16}, or with coupled superconducting resonators periodically excited by a thermal pump~\cite{hardal17}.

To realize the self-contained piston engine, we build on an established setup of a transmission line resonator terminated by a Josephson junction. This device has already served as a testbed for a variable boundary condition photon resonator~\cite{wilson10}. Specifically, the phase shift across the Josephson junction effectively realizes a mixed boundary condition, corresponding to a change in the effective length of the device. This architecture has for instance been exploited to implement a mirror moving at velocities close to the speed of light and therefore allowing for the first observation of the dynamical Casimir effect~\cite{wilson11}. Here we consider capacitive and inductive resonator-circuit coupling with corresponding half- or quarter-wave resonators and show how these two variants naturally realize a quantum piston: an object converting pressure in a cavity into rotary motion of a crank. Our particular example is unusual, in that the cavity is filled with light rather than gas, and the rotary motion is that of the time-integral of the voltage (the flux) of a superconducting island with a tunnel barrier to ground.
\section{The model}

\begin{figure}
	\centering
	\begin{overpic}[width=\columnwidth]{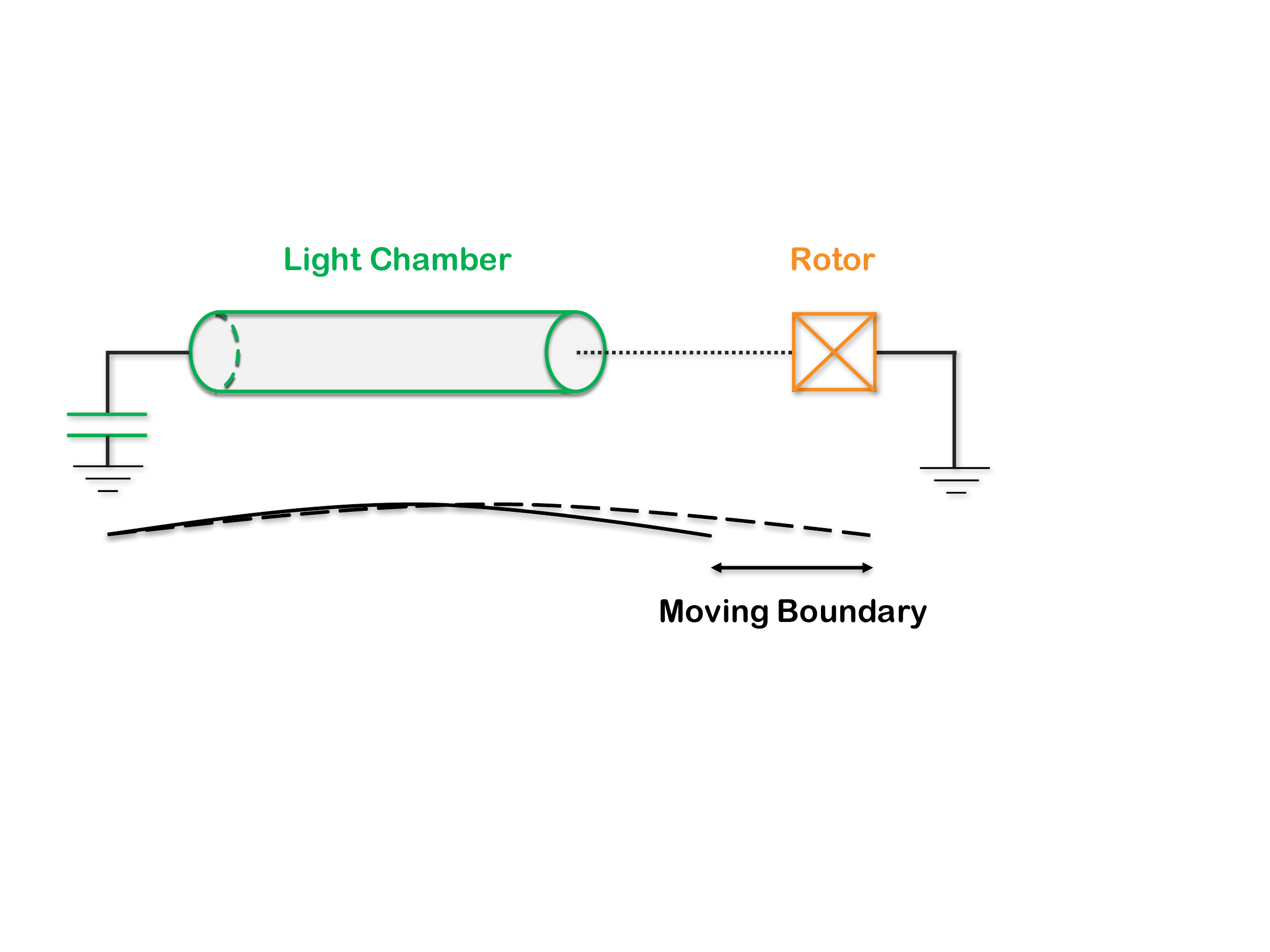}
		\put (0,38) {(a)}
	\end{overpic}
	\begin{overpic}[width=\columnwidth]{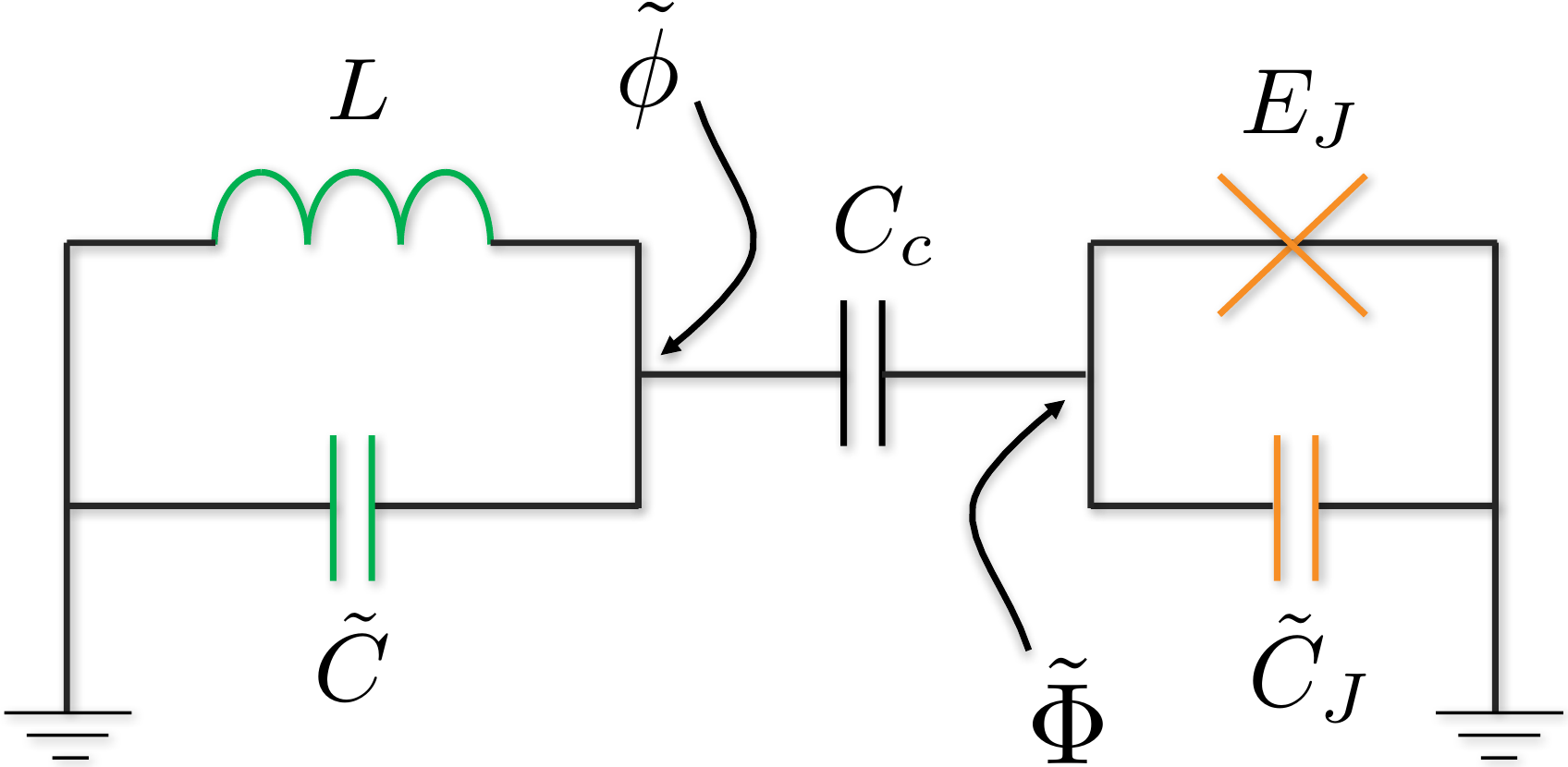}
		\put (0,45) {(b)}
	\end{overpic}
	\begin{overpic}[width=\columnwidth]{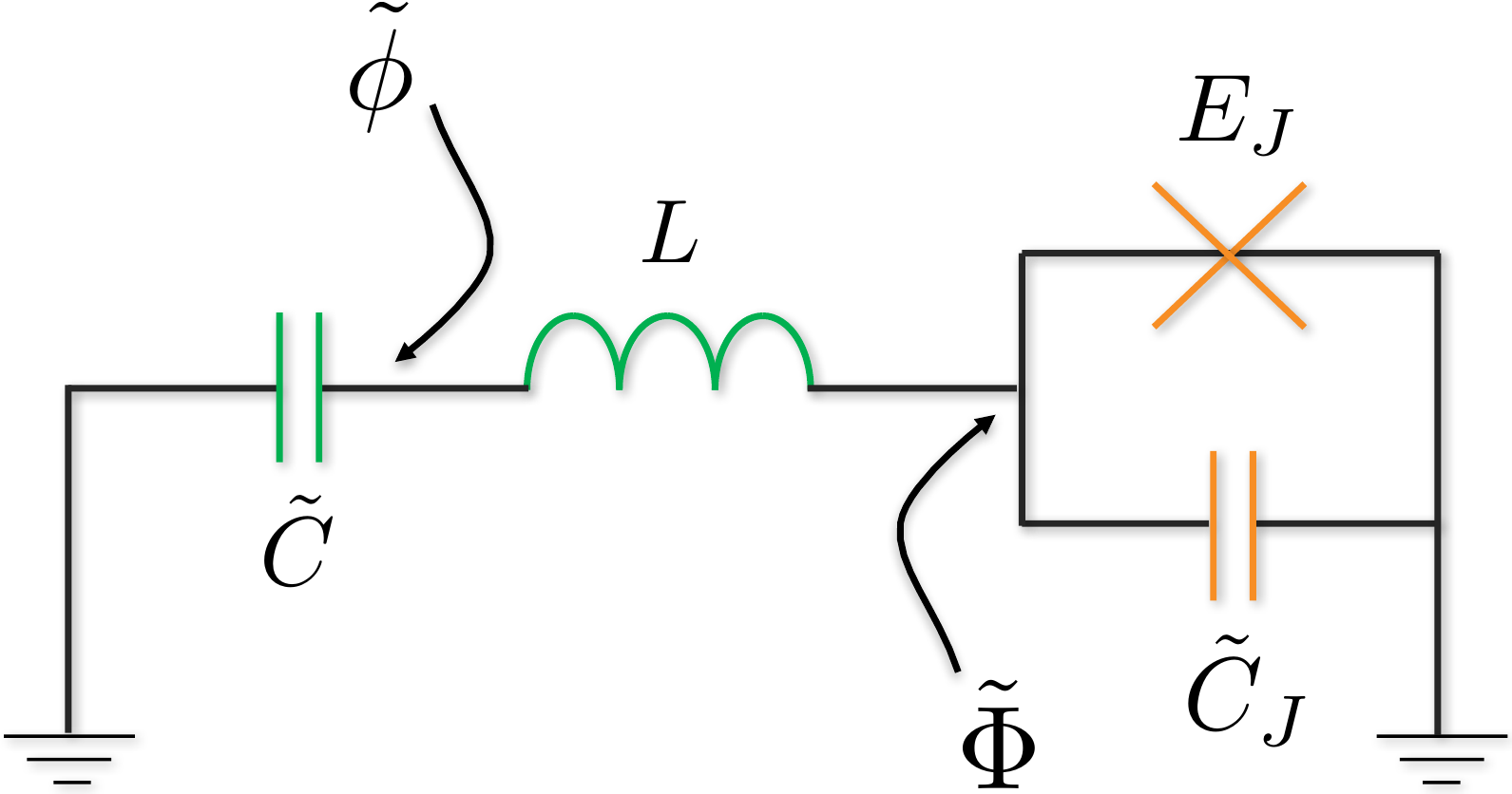}
		\put (0,45) {(c)}
	\end{overpic}
	\caption{\label{fig:design}The flux-based piston. (a) The light chamber is realized with an open boundary-condition transmission line resonator. The latter is terminated by a Josephson junction which implements the rotational degree of freedom. The dashed line represents the type of coupling. (b) Equivalent circuit diagram for a possible implementation of the device based on capacitive coupling via $C_c$. (c) Alternative implementation based on inductive coupling directly via a wire .}
\end{figure}

\emph{Design.--} The flux-based piston consists of a half-wave resonator realized by an LC circuit that is terminated by a single Cooper pair box representing a charge island~\cite{girvin09}, see Fig.\,\ref{fig:design}(a). We propose two alternative circuit designs that could essentially implement the same piston dynamics in varying parameter regimes of operation. In (b), we sketch a capacitive variant, where the parallel LC circuit representing the resonator is capacitively coupled to the Josephson loop terminating the transmission line. Panel (c) shows the second variant based on inductive galvanic coupling. The relevant bare dynamical variables in both cases are the flux quantities $\tilde\phi$ and $\tilde\Phi$ for the resonator and the loop, and measured in units of the $2\pi$-modified flux quantum $\Phi_0=\hbar/2e$. The corresponding Lagrangians are~\cite{wendin05}
\begin{eqnarray}
	\cL_{\rm cap} &=& \frac{\tilde{C}}{2} \dot{\tilde\phi}^2+\frac{\tilde{C}_J}{2} \dot{\tilde\Phi}^2+ \frac{C_c}{2}(\dot{\tilde\phi}-\dot{\tilde\Phi})^2 \nonumber \\
	&& - \frac{\tilde\phi^2}{2L}+E_J\cos \frac{\tilde\Phi}{\Phi_0} , \label{eq:Lcap} \\
	\cL_{\rm ind} &=& \frac{\tilde{C}}{2} \dot{\tilde\phi}^2 + \frac{\tilde{C}_J}{2} \dot{\tilde\Phi}^2 - \frac{(\tilde\phi - \tilde\Phi)^2}{2L} + E_J \cos \frac{\tilde\Phi}{\Phi_0} , \label{eq:Lind}
\end{eqnarray}
with $\tilde{C}, \tilde{C}_J$ the bare capacitances of resonator and Josephson loop, $C_c$  the coupling capacitance, $L$ the inductance of the resonator, and $E_J$ the Josephson energy. 
The two Lagrangians are related by a coordinate transformation and thus predict the same physical behavior of a coupled oscillator-pendulum system described by two effective phase variables $\phi$ and $\Phi$, and a dimensionless coupling parameter $\xi$, 
\begin{equation}
 \cL = \frac{1}{2} \left[ C \dot{\phi}^2 + C_J \dot{\Phi}^2 - \frac{\phi^2}{L} \right] + E_J \cos \left(\frac{\Phi + \xi \phi}{\Phi_0} \right).
\end{equation}
In the capacitive case, we achieve this by a simple displacement of the Josephson flux variable,
\begin{equation}
 \phi = \tilde\phi, \quad \Phi = \tilde\Phi - \xi \tilde\phi, \quad \xi = \frac{C_c}{C_c + \tilde{C}_J}.
\end{equation}
The effective capacitances are $C = \tilde{C} + \xi \tilde{C}_J$ and $C_J = \tilde{C}_J + C_c$.
In analogy to the rotor heat engine~\cite{rotor17}, the displaced Cooper pair box plays the role of the rotor degree of freedom. Its flux variable $\Phi$ represents the rotor angle, whereas the canonical charge variable $Q = \partial \cL / \partial \dot{\Phi}$ acts as the quantized angular momentum.
The harmonic working mode is represented by the LC circuit with its canonical quadratures $\phi$ and $q=\partial \cL/\partial \dot\phi$, and the corresponding Hamiltonian reads as
\begin{equation}
	H = \frac{q^2}{2C}+\frac{Q^2}{2C_J} + \frac{\phi^2}{2L}  - E_J \cos \left( \frac{\Phi + \xi \phi}{\Phi_0} \right).
\end{equation}
For the inductive case \eqref{eq:Lind}, we obtain the same result by switching to relative and center-of-mass coordinates,
\begin{equation}
 \phi = \tilde\Phi - \tilde\phi, \quad \Phi = \frac{\tilde{C}\tilde\phi + \tilde{C}_J \tilde\Phi}{\tilde{C} + \tilde{C}_J}, \quad \xi = \frac{\tilde{C}}{\tilde{C}+\tilde{C}_J}.
\end{equation}
Here, the center-of-mass coordinate plays the role of the rotor angle as the combined inductor-island system is isolated and should have integer charge module tunneling through the junction, the relative phase represents the harmonic amplitude of the cavity, and the effective capacitances are $C = \xi \tilde{C}_J$ and $C_J = \tilde{C}_J + \tilde{C}$.

\emph{Regime of interest.--} So far, we have derived the Hamiltonian describing the circuit designed in~Fig.\,\ref{fig:design}. It effectively describes a cavity mode with resonance frequency $\omega_0 = 1/\sqrt{LC}$ in interaction with a planar rotor whose inertia is characterized by the Josephson plasma frequency $\omega_p=\sqrt{E_J/\Phi_0^2 \tilde C_J}$. Now we demonstrate how this device corresponds to a piston transforming light pressure into rotary motion. To this end, we consider the regime of weak coupling and low cavity occupation,  $\xi^2 \avg{\phi^2}\ll\Phi_0^2$. Substituting the annihilation and creation operators of the resonator $\phi=\sqrt{\hbar/2\omega_0 C}(a^\dag+a)$ and $q=i\sqrt{\hbar\omega_0 C/2}(a^\dag-a)$, rescaling $\Phi\to\Phi/\Phi_0$ and $Q\to \Phi_0 Q$, and expanding to second order in $\xi \phi$, we then find $H=H_\text{BO}+V_\text{off-res}$ with
\begin{align}\label{eq:hbo}
	H_\text{BO}&=\hbar (\omega_0+g\cos \Phi ) \left( a^\dag a+\tfrac{1}{2} \right) + \frac{E_c}{2}Q^2-E_J\cos \Phi , \\
	V_\text{off-res}&=\frac{\hbar g}{2}\cos \Phi \left[(a^\dag)^2+a^2 \right]+\xi E_J\frac{\Phi_r}{\Phi_0}\sin \Phi \, (a^\dag+a). \nonumber 
\end{align}
Here the charging energy is defined as $E_c=1/\Phi_0^2 C_J$ and the zero-point fluctuation of the flux as $\Phi_r=\sqrt{\hbar/2\omega_0 C}$.

To complete the rotor engine analogy, we consider the limit where the cavity frequency $\omega_0$ sets the fastest time scale and exceeds by far the plasma frequency $\omega_p$ as well as the frequency modulation $g=\xi^2 \omega_p^2 \tilde C_J/2\omega_0 C\ll\omega_0$. This admits a Born-Oppenheimer-type approximation, which assumes that the slow rotor variables $\Phi$ and $Q$ are approximately constant on the fast timescale of the cavity dynamics. As a result, we can safely neglect all the off-resonant terms subsumed in $V_\text{off-res}$ that do not preserve the cavity occupation.

The piston is thus effectively described by the Hamiltonian $H_\text{BO}$, which contains a radiation pressure like-term with coupling frequency $g$ and a gravity-like pendulum potential proportional to $E_J$. The latter has its minimum at $\Phi=0$, whereas the former tends to push the rotor against that minimum potential in order to decrease the cavity resonance frequency. An inversion of the potential would occur at the critical occupation $a^\dag a\approx E_J/\hbar g$. However, this critical value is well beyond the small-occupation approximation used above and will not be relevant for what follows. Moreover, we can also ignore the vacuum correction to the gravity-like potential energy $\hbar g /2\ll E_J$, which in principle could be absorbed in a renormalization of the Josephson energy $E_J$.
\section{Thermal loading}
\emph{Autonomous operation.--} Now that we have the circuit-QED equivalent of a piston, it is time to fill up the tank and start the engine. For this, we will contact the light chamber with a hot and a cold thermal reservoir whose temperature difference shall drive the rotor motion. We aim at an integrated setup where the alternating strokes of heat extraction from the hot bath and excess heat rejection to the cold bath are synchronized with the rotation angle. This way the engine will operate autonomously, serve as its own clock, and not rely on external agents or driving fields pre-defining the sequence of engine strokes.

\begin{figure}
	\centering
	\begin{overpic}[width=\columnwidth]{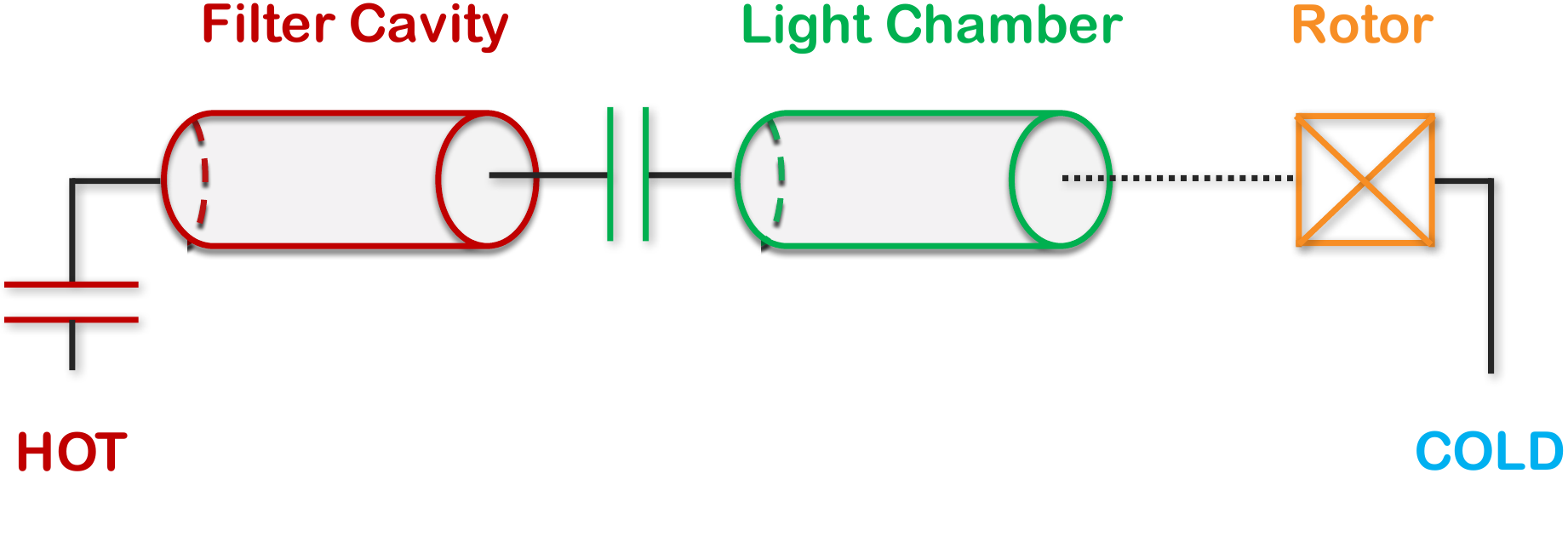}
		\put (0,35) {(a)}
	\end{overpic}
	\begin{overpic}[width=0.49\columnwidth]{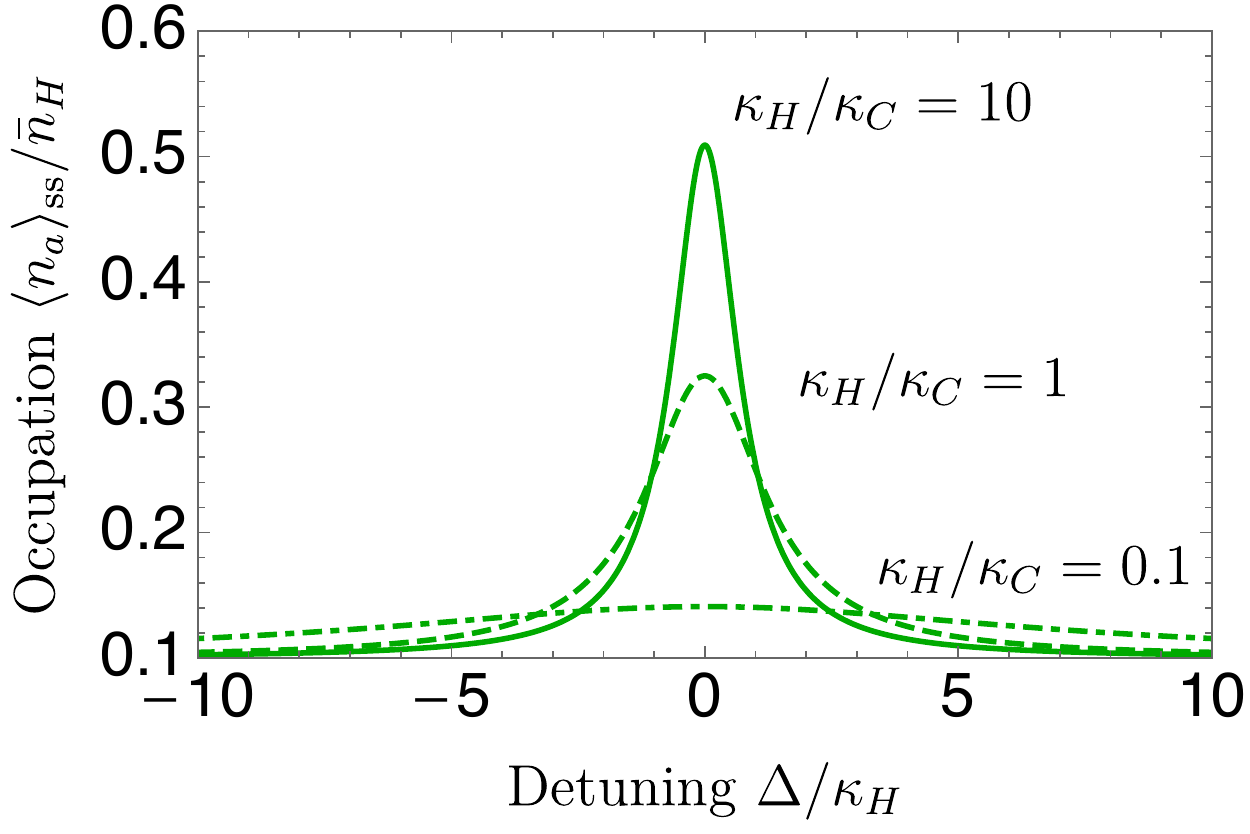}
		\put (20,54) {(b)}
	\end{overpic}
	\begin{overpic}[width=0.49\columnwidth]{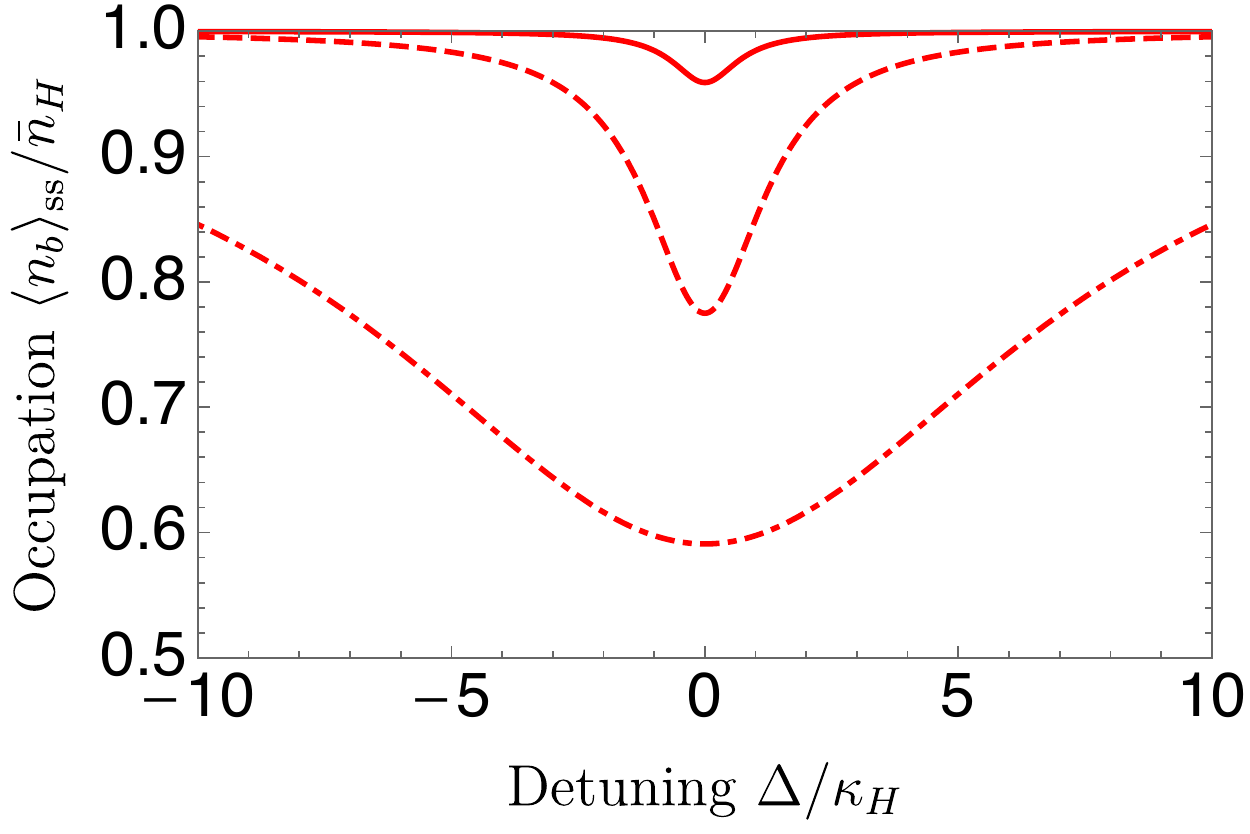}
		\put (20,54) {(c)}
	\end{overpic}
	\begin{overpic}[width=0.49\columnwidth]{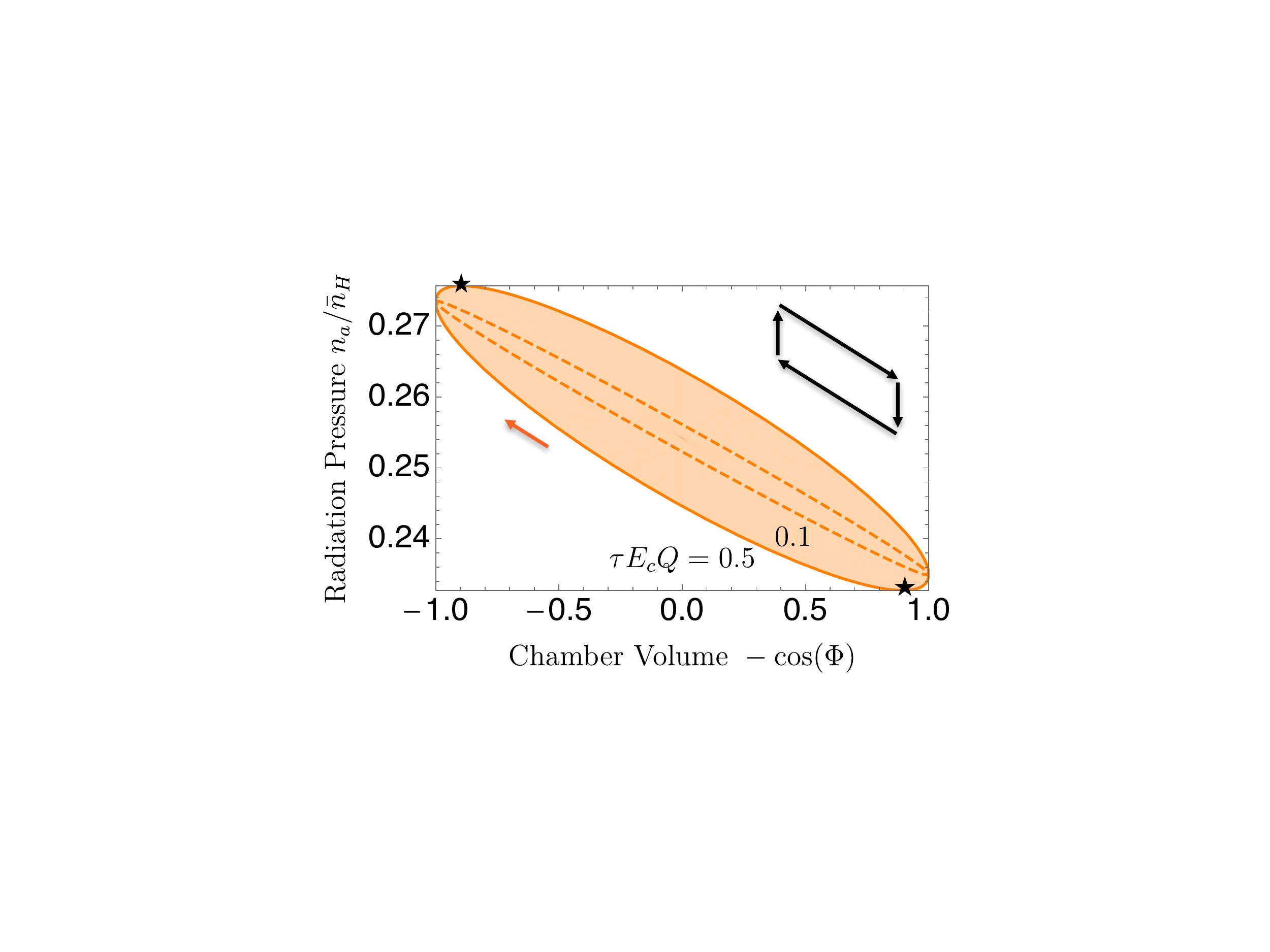}
		\put (21,19) {(d)}
	\end{overpic}
\caption{\label{fig:thLoad}Connecting the piston to the thermal resources. (a) A hot input is connected to a near-resonant filter cavity, which in turn is capacitively coupled to the light chamber of the piston. The latter is also coupled to an intrinsic cold bath, e.g.~the dilution refrigerator environment. (b) Average occupation of the light chamber in the steady-state~\eqref{eq:occSS} as a function of detuning, with $\alpha=1$ and $\bar{n}_H=10, \bar{n}_C\neq 0$. When far detuned, or for weak coupling to the hot bath $\kappa_H\ll\kappa_C$ (dot-dashed line), the cold bath dominates and the occupation stays at $\bar{n}_C$ (0.1 in units of $\bar{n}_H$). For significant coupling $\kappa_H\gtrsim\kappa_C$ (dashed and solid lines), excitations from the hot filter cavity can flow into the light chamber when the two are near resonance. This also corresponds to the regime where the average occupation of the filter cavity (c) is negligibly affected by the coupling to the system. Instead, it stays approximately thermalized with the strongly-coupled hot bath (solid line). (d) pV diagram for different non-adiabaticity, with the hypothetical Carnot cycle shown in the upper-right corner. The bare detuning $\Delta_0=-\kappa_H$ corresponds to a bare occupation $\avg{n_a}_\text{ss}(\Delta_0)\approx 0.254\,\bar{n}_H$, which is modulated by the coupling $g=0.1\, \kappa_H$ and $\kappa_H=10\, \kappa_C$. The stars mark the points of maximum and minimum light chamber occupation. They follow full compression and detente with a delay that determines the enclosed area.}
\end{figure}

In a realistic implementation of the present setup, a natural cold bath for the working mode would be provided by the system environment, as the cavity can lose photons at the rate $\kappa_C$ through either spontaneous radiation or the capacitive coupling to ground. For the modulated coupling to a hot source of thermal noise, we shall insert an additional filter cavity, as shown in Fig.\,\ref{fig:thLoad}(a). This new cavity mode $b$ with frequency $\omega_b$ shall thermalise with the hot bath at the rate $\kappa_H$ and couple to the working mode with rate $J$, via the Hamiltonian $H_{ab}=J(a^\dag b + b^\dag a)$ in the rotating-wave approximation. This ensures that transfer from this cavity to the light chamber is only significant when the two satisfy the resonance condition $\omega_b\approx\omega_0+g\cos\Phi$. The position of the rotor is therefore directly controlling the opening of the heat inlet valve, thereby implementing the necessary synchronicity that allows the engine to bypass the need for external control. In practice, the hot bath could be realized by using a resistive load on the input of the cavity or, taking advantage of the narrowband nature of the device, by driving the system with incoherent electrical noise whose variance emulates the Johnson noise of an equivalent resistor.

\emph{Equations of motion.--} To describe the engine operation, we now move to the Heisenberg-Langevin equations for the tri-partite system \cite{gardiner85}, 
\begin{align}\label{eq:eom}
	\dot b&=-(i\omega_b+\frac{\kappa_H}{2})b -iJa+\sqrt{\kappa_H}\,b_\text{in} , \nonumber\\
	\dot a&=-\left(i\omega_0+ig\cos\Phi+\frac{\kappa_C}{2}\right) a -iJb+\sqrt{\kappa_C}\,a_\text{in} , \nonumber\\
	\dot{\Phi} &= E_c Q, \quad \dot{Q} =- (E_J-\hbar g a^\dag a) \sin \Phi .
\end{align}
The input noises satisfy $\avg{a^\dag_\text{in}(t)a_\text{in}(t')}=\bar{n}_C\delta(t-t')$ and $\avg{b^\dag_\text{in}(t)b_\text{in}(t')}=\bar{n}_H\delta(t-t')$, with $\bar{n}_{H,C}$ the thermal occupation of the respective bath.

In order to see the modulation of the thermal load explicitly, let us first treat the (slow) rotor variables $\Phi$ and $Q$ as fixed parameters for the (fast) cavity dynamics. The resulting linear equations for the cavity variables are then straightforward to solve, and we obtain steady-state values for the average occupations $n_a=a^\dag a$ and $n_b=b^\dag b$ as a function of the cooperativity $\alpha=4J^2/\kappa_C\kappa_H$ and the angle-dependent detuning $\Delta(\Phi)=\omega_0+g\cos\Phi-\omega_b$,
\begin{align}\label{eq:occSS}
	\avg{n_a}_\text{ss}&=\bar{n}_C+ \frac{\alpha (\bar{n}_H-\bar{n}_C)} {\frac{4\Delta^2(\Phi)}{\kappa_H(\kappa_C+\kappa_H)}+(1+\alpha)\frac{\kappa_C+\kappa_H}{\kappa_H}} ,\nonumber\\
	\avg{n_b}_\text{ss}&=\bar{n}_H- \frac{\alpha (\bar{n}_H-\bar{n}_C)} {\frac{4\Delta^2(\Phi)}{\kappa_C(\kappa_C+\kappa_H)}+(1+\alpha)\frac{\kappa_C+\kappa_H}{\kappa_C}} .
\end{align}
These occupations are plotted against the detuning for different thermalisation rates in Fig.\,\ref{fig:thLoad}(b) and (c). We see that the $b$-mode fulfills its intended role as a frequency filter for the hot thermal source when  $\kappa_H\gg\kappa_C$. It then acts as a hot bath itself, coupling to the working mode $a$ only when the resonance condition is met. For maximum response, the angle-dependent detuning should vary on the scale of $\kappa_H$, around values close to resonance where the working mode occupation has its steepest slope. We will come back to this in a later section.

\emph{Delayed cavity reaction.--} We have shown how a filter cavity facilitates an autonomous modulation of the thermal load attached to the working mode and, thereby, can excite motion of the rotor away from its equilibrium position at $\Phi=0$. However, timing of this modulation is crucial. For an actual car piston, this is taken care of by a precise engineering of the camshaft that controls the valves opening. Here, in the quasi-static approximation \eqref{eq:occSS}, the thermal load adapts to the rotor angle as a function of the change in effective cavity length $\propto\cos \Phi$. Now this change is also what dictates the radiation pressure-like potential in ~\eqref{eq:hbo}, which means that the potential and the thermal load are modulated in the same way for positive and negative angles. Consequently, a desired net gain in directional motion can only come from corrections to the quasi-static limit due to the finite reaction time of the working mode. For the present case of a cavity mode, the reaction time is given by its inverse linewidth. It sets an effective delay for the impact of rotor-induced changes in cavity length on the cavity intensity. Note that the necessity of a finite delay is in contrast to the theoretical rotor engine model~\cite{rotor17}, which achieves the highest net gain of angular momentum per cycle when the cavity can react almost instantaneously to a change in angle.

For a qualitative estimate of the interplay between cavity and rotor dynamics, suppose that the cavity reacts to rotations with a fixed small delay $\tau$, i.e.~its occupation at any time $t$ is given by the past steady-state value
\begin{equation}\label{eq:delay}
	n_a(t)\to\avg{n_a}_\text{ss}\left(\Delta\left[\Phi(t-\tau)\right]\right) .
\end{equation}
In principle, this delay $\tau$ is both $\Phi$- and $Q$-dependent, but we shall keep it fixed for now out of simplicity.

To lowest order in $\tau$ and in the modulation of the detuning $g\ll\kappa_H$, we expand $\Phi(t-\tau)\approx \Phi(t)-\tau\dot\Phi(t)$ and %
\begin{equation}
	\avg{n_a}_\text{ss}\left(\Delta\left[\Phi\right]\right)\approx \avg{n_a}_\text{ss}\left(\Delta_0\right)+g\cos(\Phi) \frac{\partial \avg{n_a}_\text{ss}}{\partial\Delta}\Big|_{\Delta_0} ,
\end{equation}
where $\Delta_0=\omega_0-\omega_b$ is independent of $\Phi$. 
Inserting this into the equation \eqref{eq:eom} for the rotor momentum, we obtain
\begin{align}\label{eq:dQqual}
	\dot Q&=-\Big[E_J-\hbar g \avg{n_a}_\text{ss}\left(\Delta\left[\Phi(t-\tau)\right]\right)\Big] \sin\Phi \nonumber\\
	&\approx C_1\sin \Phi  + C_2\sin \Phi \cos \Phi +C_3 Q \sin^2 \Phi ,
\end{align}
with the constants $C_1=\hbar g\avg{n_a}_\text{ss}(\Delta_0)-E_J$, $C_2=\hbar g^2\frac{\partial \avg{n_a}_\text{ss}}{\partial\Delta}\Big|_{\Delta_0}$ and $C_3=\tau E_c C_2$. 
Here, the first two terms describe the quasi-static case of a cavity reaction without delay; they average to zero over a $2\pi$ rotation period. A net buildup of directional rotation can only come from the delay itself, through the $\tau$-proportional third term, if $C_2>0$. This requires us to operate with a blue-detuned filter cavity, $\Delta_0 < 0$, as opposed to a red-detuned one that would lead to friction.

In thermodynamic terms, the gain process then constitutes a Carnot-type engine cycle with four strokes. At the point of smallest volume, $\Phi=0$, the chamber cavity is closest to resonance and fills with thermal photons from the hot filter cavity (pressure increases). As the rotor starts moving, the chamber expands (volume increases) and moves away from resonance. Consequently, excess photons leak out through the continual thermal contact into the cold bath (pressure decreases). Driven by inertia, the rotor continues its rotation cycle and compresses the chamber (volume decreases) again. The strokes are shown in a pV diagram in Fig.\,\ref{fig:thLoad}(d). 
Contrary to the ideal Carnot cycle, the engine is in contact with the cold bath at all times, while the heat inlet from the hot bath continuously opens and closes as a function of the chamber volume.

The enclosed area in the diagram determines the net work per cycle the cavity performs to accelerate the piston $\mathcal{W}_\text{cyc}=\pi \tau E_c Q C_2$. This transient gain requires no external control, but crucially depends on the rotation frequency and the finite time it takes the chamber cavity to build up and release its radiation pressure. Each cycle thus varies in both its duration and the area covered by the pV diagram. The generated work accumulates in the form of directed piston rotation, which could be extracted by means of an external load (provided its pull does not exceed the thermal gain) \cite{stickler2017,seah2018,niels18}. Note however that such an electrical ``engine'' is restricted to low efficiencies. Indeed, similar to the engine of Ref.~\cite{rotor17}, the radiation pressure like-term each thermal photon contributes to the work output is of order $C_2\sim g$, which is much smaller than its energy $g\ll \omega_0$.
\section{Piston in action}
We will now show that the piston works as intended and generates a thermally-driven net gain of rotation. To this end, we shall simulate the stochastic equations of motion~\eqref{eq:eom} in the classical regime, fully accounting for the noise inputs of the two baths.

\emph{Eliminating the filter cavity.--} 
In the relevant regime of fast thermalisation with the hot bath, $\kappa_H\gg J,\kappa_C,E_c Q$, we can simplify the description by adiabatically eliminating mode $b$. This reduces the filter cavity to its effective role of a hot bath whose coupling rate is modulated by the rotor's position via $\Delta(\Phi)$. Moreover, the rotor only reacts to changes in the light chamber's occupation $n_a$, therefore we do not need to track its phase evolution $\arg (a)$. Using Ito calculus~\cite{gardiner04} and noting that the evolution of a two-dimensional Orstein-Uhlenbeck process tends to white noise in the limit of vanishing correlation time $\kappa_H^{-1}$, we then obtain for the light chamber dynamics
\begin{equation}\label{eq:dna}
	\rmd n_a = -\kappa(\Phi) [n_a-\bar{n}(\Phi)]\rmd t + \sqrt{2\kappa(\Phi) \bar{n}(\Phi)n_a} \rmd W ,
\end{equation}
with $W$ a Wiener process. The two complex variables $a$ and $b$ and their associated noise terms are thus reduced to a single real-valued random variable, the chamber occupation $n_a$, coupled to the rotor. The physics described by this equation is that of thermalization to a $\Phi$-dependent value
\begin{equation}
	\bar{n}(\Phi)=\lim_{\kappa_H\gg\kappa_C}\avg{n_a}_\text{ss}(\Phi)=\frac{\bar{n}_C+f_H(\Phi)\bar{n}_H}{1+f_H(\Phi)} ,
\end{equation}
at a $\Phi$-dependent rate $\kappa(\Phi)=\kappa_C[1+f_H(\Phi)]$. The angle-dependent function describes the hot thermal contact to $\bar{n}_H$, mediated by the filter cavity mode $b$, 
\begin{equation}
	f_H(\Phi)=\frac{\alpha}{1+\frac{4\Delta^2(\Phi)}{\kappa_H^2}} .
\end{equation}
In the rest of this manuscript, we consider the case of unit cooperativity $\alpha=1$ such that $0\leq f_H(\Phi)\leq 1$ and $1\leq\kappa(\Phi)/\kappa_C\leq 2$.

\emph{Simulations.--}
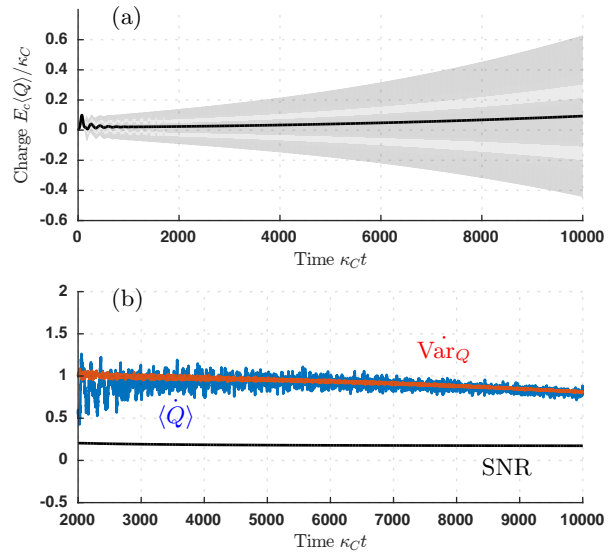
\begin{figure}
	\centering
	\begin{overpic}[width=\columnwidth]{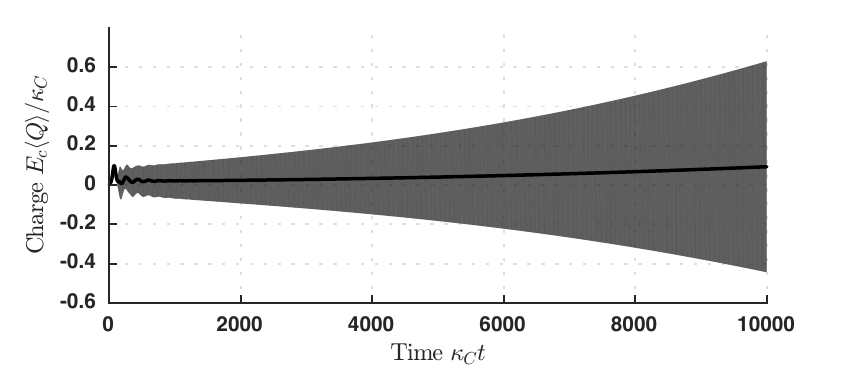}
		\put (18,38) {(a)}
	\end{overpic}
	\begin{overpic}[width=\columnwidth]{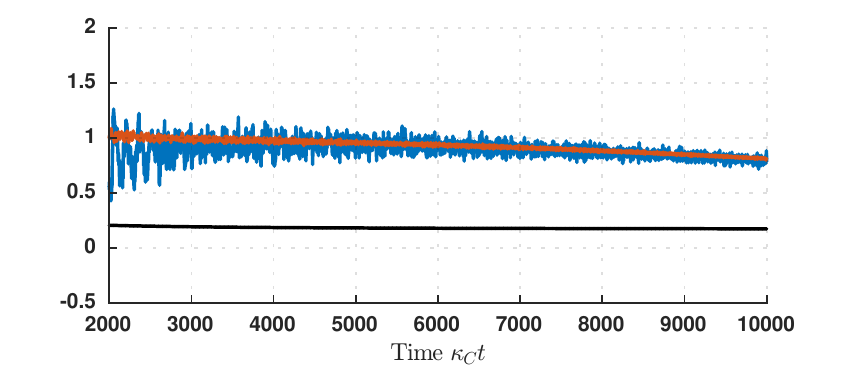}
		\put (18,38) {(b)}
		\put (25,20) {\color{blue}$\dot{\avg{Q}}$}
		\put (65,30) {\color{red}$\dot{\text{Var}_Q}$}
		\put (75,12) {SNR}
	\end{overpic}
\caption{\label{fig:sim}Simulations of the piston under thermal loading. (a) Starting at $\Phi=-0.95\pi$ and at rest, a net charge is accumulated on average. The grey area covers one standard deviation. (b) The rate of increase of the average (blue) and variance (red) are normalized respectively to the analytical prediction of Eqs.\,\eqref{eq:gainFR} and \eqref{eq:varFR}. The agreement (\emph{i.e} a value of 1) is not valid anymore when second- and higher-order corrections come into play. Here, this corresponds to the regime where $E_c^2\avg{Q^2}/\kappa_C^2\approx 0.1$, which is reached for $\kappa_C t\approx 6770$. For reference, the signal-to-noise ratio is shown to be approximately constant. The parameters for the simulation are $E_J E_c=\hbar E_cg\bar{n}_H=0.004\kappa_C^2$ and $\bar{n}_H=100\bar{n}_C$.}
\end{figure}
An exemplary simulation result is shown in Fig.\,\ref{fig:sim}(a). 
It starts at rest from an angle $\Phi = -0.95 \pi$ close to the maximum of the pendulum potential $-E_J\cos(\Phi)$. 
The gain dynamics is present, and it causes the Cooper pair box to accumulate enough net average charge (angular momentum) to overcome the potential after one pendular swing and continue rotating clockwise. Without the gain, the system would remain in swinging pendulum motion around $\Phi=0$. On average, the piston behaves as desired, but with a significant amount of noise on top of the dynamics of interest. This is inherent to the design, since the gain is a first order effect in the delayed cavity reaction, whereas the momentum diffusion is directly caused by the thermal noise input without delay. Yet, we note that once the engine has entered the regime of rotation, the signal-to-noise ratio $\avg{Q}/\sqrt{\text{Var}_Q}$ remains approximately constant (see the black line in Fig.\,\ref{fig:sim}(b)).

\emph{Analytical model.--}
Next we derive an approximate expression for the average gain in the working regime of unbounded rotation. Rather than introducing a fixed delay parameter $\tau$ by hand, as done in~Eq.\,\eqref{eq:delay}, we now expand the actual equation of motion~\eqref{eq:dna} as
\begin{align} \label{eq:na_1stDelay}
	n_a&=\bar{n}(\Phi)-\bar{n}^\prime(\Phi) E_c Q/\kappa (\Phi)+e^{-\kappa(\Phi)t}\varepsilon, \\
	\rmd \varepsilon &\approx e^{\kappa(\Phi)t}\sqrt{2\kappa(\Phi)\bar{n}(\Phi)n_a}\rmd W ,\nonumber
\end{align}
Here the velocity- or $Q$-dependent term will be responsible for the gain.
The expansion is valid as long as the rotation frequency $E_c Q$ is approximately constant on the thermalisation time scale $1/\kappa(\Phi)$ \footnote{In this regime, the rotor angle $\Phi(t)$ and functions thereof will then remain non-anticipating for any Wiener process $W_{t^\prime}$ where $t^\prime$ is close to $t$ on a timescale $\kappa(\Phi)^{-1}$.}, i.e.
\begin{equation}
	E_c E_J, \hbar g E_c, E_c^2 Q^2\ll \kappa(\Phi)^2 .
\end{equation}
The last $Q$-dependent requirement identifies an ideal regime of operation. If the rotor is much faster, the cavity can no longer follow and its $\Phi$-dependent reaction averages out. A very slow rotor, on the other hand, would only cause little gain as the cavity could react to the changing angle on time. Heuristically, the sweet spot for achieving maximum output per cycle thus corresponds to the critical regime of rotation frequencies $E_c Q \approx 0.1\kappa(\Phi)$. This is in accordance with our simulations, where the gain starts to deteriorate at greater frequencies (see Fig.~\ref{fig:sim}(b)).

Substituting the expansion \eqref{eq:na_1stDelay} into the equation \eqref{eq:eom} for the rotor momentum, we find
\begin{equation}\label{eq:dQgain}
	\rmd Q = \left[ g_\text{odd}(\Phi)+g_\text{non-ant.}(\Phi)e^{-\kappa(\Phi)t}\varepsilon +\chi(\Phi)Q \right]\rmd t .
\end{equation}
Here the first term is an odd function of the angle, which yields no net effect over a cycle, and the second one is a noise term that averages to zero. The last term, which arises from the delay in the thermalization, is the one of interest; it describes momentum gain (or dissipation) at the rate
\begin{align}
	\chi(\Phi)&=-\frac{\hbar gE_c}{\kappa(\Phi)}\bar{n}^\prime(\Phi) \sin\Phi \\
	&=-\frac{\hbar gE_c}{\kappa(\Phi)}\left(\bar{n}_H-\bar{n}_C\right)8\alpha \sin^2 \Phi \frac{g\Delta(\Phi)/\kappa_H^2}{\left[1+\alpha+ \frac{4 \Delta(\Phi)^2}{\kappa_H^2}\right]^2} .\nonumber
\end{align}
This result is similar to the qualitative description of Eq.\,\eqref{eq:dQqual}, but with a physical expression for the gain term that does not depend on an adhoc constant $\tau$. As found out previously, the filter cavity needs to be blue-detuned with respect to the light chamber for obtaining gain as opposed to friction. In particular, the maximum gain at the angle of strongest radiation pressure is obtained when setting $\Delta_0=-\sqrt{(1+\alpha)/12}\approx-0.4\kappa_H$. 
Friction is avoided as long as the frequency modulation by the rotor does not change the sign of the detuning, $g\leq \Delta_0$.

In the free-rotation limit~\cite{rotor17}, once the rotor has overcome its initial stage of pendulum oscillations and rotates at a steadily but slowly increasing frequency, we can approximate the net speedup per cycle by the angle-averaged formula
\begin{equation}\label{eq:gainFR}
	\dot{\avg{Q}}=\chi \avg{Q} , \quad \chi=\frac{1}{2\pi}\int_0^{2\pi}\chi(\Phi)\rmd \Phi.
\end{equation}
It can be understood as an accumulation of static charge caused by a net current of photons through the circuit from the hot to the cold bath. We omit here the exact expression for $\chi$, which is rather lengthy and uninformative. Instead, we compare it directly to the numerical simulation results in Fig.\,\ref{fig:sim}(b). We find good agreement up to times when the piston starts being too fast for the light chamber to follow. This is where our first-order expansion fails, high-order delay corrections become significant, and the gain decreases.

From the first order expansion \eqref{eq:na_1stDelay}, we also obtain an analytical expression for the growth of the charge variance $\text{Var}_Q=\avg{Q^2}-\avg{Q}^2$. The derivation is more complicated as it involves higher moments of the noise term $\varepsilon$. However, the result is greatly simplified when considering $2\pi$-averages in the free-rotation regime, 
\begin{equation}\label{eq:varFR}
	\dot{\text{Var}_Q}=2\chi \text{Var}_Q+ \frac{\hbar^2g^2}{\pi}\int_0^{2\pi}\frac{\bar{n}^2(\Phi) \sin^2 \Phi}{\kappa(\Phi)}\rmd \Phi .
\end{equation}
Fig.\,\ref{fig:sim}(b) shows that this is also in good agreement with the simulation, and that the corresponding signal-to-noise ratio is indeed approximately constant.
\section{Conclusion}
We have proposed an autonomous flux piston based on the circuit QED architecture. The built-in synchronicity allows the piston to extract a net positive charge when thermally loaded, without the action of an external agent. This is achieved by modulating the resonance condition between the light chamber and a filtered hot bath via the engine's rotational degree of freedom. While the current design still suffers from the very noisy output generated by the piston, it constitutes a first proof-of-principle for a fully autonomous rotor heat engine that can in principle operate in the quantum regime and therefore address questions related to the worthiness of non-thermal resources. Of particular note is the relative ease of implementing squeezed reservoirs using circuit QED, and thus such explorations are in the immediate grasp of experiments. Moreover, it may also prove useful as a testbed to explore more general concepts, such as the cost of keeping time~\cite{erker17}.

\begin{acknowledgments}
We thank V.~Manuchuryan and P.P.~Hofer for helpful feedback. JMT thanks CQT for their kind hosting during his stay. This research is supported by the Singapore Ministry of Education through the Academic Research Fund Tier 3 (Grant No. MOE2012-T3-1-009); by the National Research Foundation, Prime Minister’s Office, Singapore, through the Competitive Research Programme (Award No. NRF-CRP12-2013-03); and by both above-mentioned source, under the Research Centres of Excellence programme. In addition, this work was financially supported by the Swiss SNF and the NCCR Quantum Science and Technology.
\end{acknowledgments}

\bibliography{pistonBib}

\end{document}